\documentclass[%
 reprint,
 amsmath,amssymb,
 aps,
]{revtex4-2}
\usepackage{comment}
\usepackage{xcolor}
\usepackage{graphicx}
\usepackage{dcolumn}
\usepackage{bm}

\usepackage{enumitem}
\newcommand{\mR}{\mathbb R}

\newcommand{\black}{\color{black}}

\begin{document}

\preprint{APS/123-QED}

\title{
Inferring potential landscapes: \\A Schr\"odinger bridge approach to Maximum Caliber}

\author{Olga Movilla Miangolarra }
 
\author{Asmaa Eldesoukey}%
\author{Tryphon T. Georgiou}

\affiliation{%
Department of Mechanical and Aerospace Engineering, University of California, Irvine, CA 92697, USA}%
\email{Emails: omovilla@uci.edu, aeldesou@uci.edu, tryphon@uci.edu.\\
This research has been supported in part by the AFOSR under FA9550-23-1-0096, and the ARO under W911NF-22-1-0292.}



\date{\today}

\begin{abstract}
 Schr\"odinger bridges have emerged as an enabling framework for unveiling the stochastic dynamics of systems based on marginal observations at different points in time. The terminology ``bridge'' refers to a probability law that suitably interpolates such marginals. The theory plays a pivotal role in a variety of contemporary developments in machine learning, stochastic control, thermodynamics, and biology, to name a few, impacting disciplines such as single-cell genomics, meteorology, and robotics. In this work, we extend Schr\"odinger's paradigm of bridges to account for integral constraints along paths, in a way akin to  Maximum Caliber -- a Maximum Entropy principle applied in a dynamic context. The Maximum Caliber principle has proven useful to infer the dynamics of complex systems e.g., that model gene circuits and protein folding.
We unify these two problems via a maximum likelihood formulation to reconcile stochastic dynamics with ensemble-path data. A variety of data types can be encompassed, ranging from distribution moments to average currents along paths. 
The framework 
enables inference of time-varying potential landscapes that drive the process. The resulting forces can be interpreted as the optimal control that drives the system in a way that abides by specified integral constraints. This, in turn, relates to a similarly constrained Optimal Mass Transport problem in the zero-noise limit.
Analogous results are presented in a discrete-time, discrete-space setting and specialized to steady-state dynamics. We finish by illustrating the practical applicability of the framework through paradigmatic examples, such as that of bit erasure or protein folding.
In doing so, we highlight the strengths of the proposed framework, namely, the generality of the theory, the ease of computation, and the ability to interpret results in terms of system dynamics. This is in contrast to Maximum-Caliber problems where the focus is typically on updating a probability law on paths.

\end{abstract}

\maketitle

\section{Introduction}
In 1931 Schr\"{o}dinger posed the following problem \cite{Sch31,Sch32}:
    Suppose we start with an initial distribution $p_0$ of particles that obey known stochastically driven dynamics. These specify the distribution $p_t$, at subsequent times $t\in[0,T]$, via the corresponding Fokker-Planck equation, assuming an overwhelmingly large population.
    However, for a finite population of, say $N$ particles, the empirical distribution at time $t$ may differ from the one prescribed for typical events. Indeed, when the number of particles is finite, rare events are of interest.
    Thus, assume we have an ``auxiliary observer'' \cite{Sch31,chetrite2021schrodinger} who observes the position of the particles at times $t\in[0,T]$, without however reporting us the results. At time $T$, we measure the empirical distribution of particles $p^{\rm exp}_{T}$, which does not match the prediction $p_{T}$ obtained by solving our dynamics. Clearly, a rare event took place. That is, something unlikely happened, but what? For such instances, it is of interest to determine what probabilistic inferences we can draw from the two observations (at time $0$ and $T$) regarding possible intervening observations by the auxiliary observer.
In other words, we are interested in the most likely path that the particles may have taken in their flow from $p_0$ to $p_{T}$, given the dynamics (\emph{prior} evolution). 

This problem has been studied ever since under the name of Schr\"odinger bridge (SB)  \cite{wakolbinger1931to1991,leonard2013survey}. It turns out that the solution can be traced forward (as this subject was not in existence at the time Schr\"odinger raised and answered the question) to a field of Probability Theory known as Large Deviations Theory and, in particular, to Sanov's Theorem \cite{dembo2009large}.
Sanov's Theorem states that the probability of drawing an atypical distribution for a finite collection of $N$ particles, asymptotically, as $N\to\infty$, decays exponentially to zero as $\sim e^{-H}$, where $H$ is the relative entropy between the atypical observed distribution and the typical one (prior). Let us denote the atypical distribution by $P$. This is a distribution on the space of continuous paths, unknown, apart from its marginal distributions at times $0$ and $T$. Therefore, Schr\"odinger's dictum is to find the most likely atypical distribution, given the prior distribution $Q$ as specified by the dynamics of our stochastic system.  That is, Schr\"odinger postulated (and solved) the maximum likelihood problem to find
$$
\min_{P} H(P, Q),
$$
where $P$ is such that its marginal distributions at times $0$ and $T$ match those that have been observed empirically, i.e., $p_0=p_0^{\rm exp}$ and $p_{T}=p_T^{\rm exp}$.

The Schr\"odinger bridge problem has had a multitude of applications, ranging from economics to biology, impacting disciplines such as single-cell dynamics \cite{schiebinger2019optimal}, meteorology \cite{fisher2009data}, or robotics \cite{elamvazhuthi2019mean}.
Nevertheless, the type of measurements that have been considered before do not account for constraints and data that may originate in a variety of physical experiments, such as those pertaining to currents, moments of distributions, or occupation probability of different states. 
This type of measurements is typically accounted for in a principle known as Maximum Caliber.

Specifically, the Maximum Caliber principle (or MaxCal for short), introduced by Jaynes in 1980~\cite{jaynes1980minimum}, extends the principle of Maximum Entropy to a dynamic context~\cite{presse2013principles}. Following a similar rationale, it maximizes entropy over trajectories, taking into account constraints based on specific information gathered from the system~\cite{ghosh2020maximum}. This approach allows for the incorporation of various types of new data, often involving currents, while maintaining maximum uncertainty about the remainder of the system. The Maximum Caliber principle has proven highly effective as an inference method, particularly in the context of complex systems with a small number of particles \cite{dixit2018perspective}, such as gene circuits \cite{firman2017building}, protein folding~\cite{wan2016maximum}, bird flocking \cite{cavagna2014flocking} or network traffic \cite{ghosh2020maximum}. Notably, it has successfully determined reaction rates in biomolecular simulations, including peptides \cite{wan2016maximum} and protein-protein interactions \cite{zhou2017bridging}.

Motivated by this, and in particular, by the connection of  Sch\"odinger's paradigm to 
the principle of Maximum Caliber, we extend the existing theory of Schr\"odinger bridges to account for diverse measurements. 
Specifically, the work is organized as follows. After an introduction to classical Sch\"odinger bridges in Section \ref{sec:CSB}, we consider the Maximum Caliber problem from the perspective of Schr\"odinger bridges (Section \ref{sec:MaxCalSB}). Therein, we tackle two problems: first, a standard Schr\"odinger bridge problem with an extra ensemble-path constraint, and then the standard Maximum Caliber problem with free initial and final marginals, which we also specialize to steady-state.
Section \ref{sec:discrete}, deals with the counterpart of the theory in the setting of 
discrete-time discrete-space Markov chains.
Finally, Section
\ref{sec:examples} illustrates the developed framework with examples that highlight the relevance and generality of the theory in inferring potential landscapes.

\section{Classical Schr\"odinger bridges}\label{sec:CSB}

Consider two probability laws $P$ and $Q$ on the space of continuous functions on $[0,T]$, denoted by $\Omega=\mathcal C([0,T])$,  and thought of as a space of paths. We say $P$ is absolutely continuous with respect to $Q$, denoted by $P\ll Q$ if $P$ has measure zero on all sets in which $Q$ has measure zero. In that case, we can define the relative entropy between $P$ and $Q$ as
$$
H(P,Q)= \int_\Omega dP\log\frac{dP}{dQ}.
$$
Let $Q$ represent a given (prior) probability law, and $p_t$ denote the one-time marginals of $P$. The Schr\"odinger bridge problem can be formalized as follows:
\begin{align}\label{eq:SB}
   & \min_{P:P\ll Q} \ H(P,Q)\\\nonumber
 & {\rm s.t.} \ \int_\Omega dP=1,\   p_0=p_0^{\rm exp},\ 
    p_T=p_T^{\rm exp},
\end{align}
this is, find the most likely law $P$ that satisfies the experimentally observed marginals $p_0^{\rm exp},\,p_T^{\rm exp}.$

\subsection{Laws on paths}
 We formally seek a re-weighting of the (prior) probability law $Q$ on paths so that the (posterior) $P$ satisfies the constraints. To do so, let
$\{X_t\}$ denote the canonical process $X_t=\omega_t$, with $\omega\in\Omega$.
The Lagrangian for the optimization problem in this 
path space representation can be written as 
\begin{align*}
\mathcal L&= \int_\Omega \Big(\log\frac{dP}{dQ}+\gamma+\mu(X_0)+ \eta(X_T) \Big)dP\\
&= \int_\Omega \Big(\log\Lambda +\gamma+\mu(X_0)+ \eta(X_T) \Big)\Lambda dQ,
\end{align*}
where $\Lambda(\omega)=\frac{dP}{dQ}(\omega)$ is the Radon-Nikodym derivative of $P$ with respect to $Q$, and $\gamma, \ \mu(X_0)$ and $\eta(X_T)$ are Lagrange multipliers that allow imposing the constraints on $P$.
The first variation of the Lagrangian with respect to  $\Lambda$ is given by
$$
\delta \mathcal L=\int_\Omega \Big(\log\Lambda 
+\gamma+1+\mu(X_0)+ \eta(X_T) \Big)\delta dQ,
$$
for $\delta(\omega)$ a function on $\Omega$. Setting this to zero for all $\delta$, we obtain the first-order optimality condition
$$
\Lambda^\star 
=\frac{e^{-\mu(X_0)}e^{- \eta(X_T)}}{\mathcal Z},
$$
where $\mathcal Z=e^{\gamma+1}$ is the normalization constant or partition function. Hence, the optimal law is
\begin{equation}\label{eq:fgQ}
    P^\star=f(X_0)g(X_T) Q,
\end{equation}
where $f(X_0)\propto e^{-\mu(X_0)}$ and $g(X_T)\propto e^{-\eta(X_T)}$, up to the constant normalization factor that we have absorbed. This is the standard structure of the Schr\"odinger bridge problem \cite{leonard2013survey,chen2021stochastic}, which has the property of being Markovian whenever the prior $Q$ is Markovian \cite[Prop 2.10]{leonard2013survey}. It remains to characterize the functions $f(X_0)$ and $g(X_T)$ by imposing the endpoint constraints. Instead of doing so directly, it is informative to consider a dynamical description (following \cite{chen2016relation}), for which we restrict ourselves to diffusion processes.


\subsection{A stochastic control perspective}
Consider 
that $Q$ is the law of
a diffusion process that obeys the It\^o stochastic differential equation
\begin{align}\label{eq:p-dyn}
    dX_t=\alpha(t,X_t)dt+\sigma(t,X_t) dW_t,\ X_0\sim q_0,
\end{align}
where $X_t\in\mathbb R^n$,  $W_t\in\mathbb R^n$ is an
$n$-dimensional Brownian motion, $\alpha(t,X_t)\in\mathbb R^n$ is the drift of the prior dynamics and $\sigma(t,X_t)>0$ the diffusion coefficient. For notational simplicity we assume throughout that $\sigma$ is scalar; all results can be generalized to the case of a matricial $\sigma$ in a straightforward manner \footnote{Simply substitute $\Delta(\sigma(t,x)^2(\cdot))$ and $\sigma(t,x)^2\Delta(\cdot)$  by 
$$
\sum_{i,j=1}^n \partial_{x_i}\partial_{x_j}D_{ij}(\cdot) \ \mbox{ and }\ \sum_{i,j=1}^n D_{ij}\partial_{x_i}\partial_{x_j}(\cdot),
$$
respectively,
where $D=\sigma\sigma'$}.

The probability density function corresponding to \eqref{eq:p-dyn} obeys the Fokker-Planck equation 
     \begin{align}\label{eq:FP1}
\!\! \partial_tq(t,x)\!=\!-\nabla\cdot(\alpha(t,x)q(t,x))\!+\!\frac12\Delta (\sigma(t,x)^2q(t,x)),
\end{align}
with $ q(0,x)=q_0(x)$. 
 Since the prior $Q$ is Markovian, the Schr\"odinger bridge $P^\star$ in \eqref{eq:fgQ} is also Markovian and absolutely continuous with respect to $Q$. It follows then that $P^\star$ is the law of a stochastic process that obeys
\begin{align}\label{eq:Xstar}
dX_t=\beta(t,X_t)dt+\sigma(t,X_t)dW_t,\ X_0\sim p_0^{\rm exp},
\end{align}
for a new drift $\beta(t,X_t)\in\mathbb R^n$, and thus, the (one-time) probability density obeys the Fokker-Planck equation

\begin{align}\label{eq:FP2}
\!\! \partial_tp(t,x)\!=\!-\nabla\cdot(\beta(t,x)p(t,x))\!+\!\frac12\Delta (\sigma(t,x)^2p(t,x)),
\end{align}
with $p(0,x)=p_0^{\rm exp}(x)$.

Indeed, for the law of \eqref{eq:Xstar} to be absolutely continuous with respect to the law of \eqref{eq:p-dyn}, it is necessary that the diffusion coefficients are the same. Then, by Girsanov's theorem, the Radon-Nikodym derivative $\Lambda^\star=dP^\star/dQ$
can be explicitly written as \cite[Section 3.5]{karatzas2012brownian}
$$
\Lambda^\star\!=\frac{p^{\rm exp}_0\!\!}{q_0}e^{\textstyle\int_0^T\!\frac{\|\beta(t,X_t)-\alpha(t,X_t)\|^2}{2\sigma(t,X_t)^2}{\scriptstyle dt\,+}\!\int_0^T\! \frac{\beta(t,X_t)'-\alpha(t,X_t)'}{\sigma(t,X_t)}{\scriptstyle dW_t}},
$$
where $'$ denotes transpose.
Taking the logarithm and expectation,
 Schr\"odinger's problem \eqref{eq:SB} can be recast as
\begin{align}\label{eq:dyn-SB}
   & \min_{p,\beta} \frac{1}{2}\int_0^T\int_{\mR^n}\frac{\|\beta(t,x)-\alpha(t,x)\|^2}{\sigma(t,x)^2}p(t,x)dxdt\\[3pt]\nonumber
 & \quad \mbox{such that }  p(0,\cdot)=p_0^{\rm exp},\ 
    p(T,\cdot)=p_T^{\rm exp},\\
    &\quad \mbox{and } \,\partial_tp=-\nabla\cdot(\beta p)+\frac12\Delta (\sigma^2p).\nonumber
\end{align}
To derive \eqref{eq:dyn-SB} we have used firstly the fact that the term $\int_{\mR^n} p_0^{\rm exp}\log\frac{p_0^{\rm exp}}{q_0}dx$ is fixed by the boundary conditions, and thus can be omitted from the optimization problem, and secondly, the fact that the expectation of the martingale term $\int_0^T \frac{\beta(t,X_t)'-\alpha(t,X_t)'}{\sigma(t,X_t)}{dW_t}$ vanishes.

For completeness of exposition, we note that the optimal probability density $p(t,x)$, at any time $t\in[0,T]$, is known to factor as
\begin{equation}\label{eq:factored}
    p(t,x)=\varphi(t,x)\hat\varphi(t,x),
\end{equation}
where $\varphi$ and $\hat\varphi$ solve the Schr\"odinger system
of PDEs
\begin{subequations}\label{eq:Schrod-sys}
\begin{align}\label{eq:varphi}
\partial_t\varphi(t,x)&=-\alpha(t,x)'\nabla\varphi(t,x)-\frac{\sigma(t,x)^2}{2}\Delta \varphi(t,x),
    \\\label{eq:hatvarphi}
\partial_t\hat\varphi(t,x)\,\,\,&\!\!\!\!\!=\!-\nabla\!\cdot\!(\alpha(t,x)   \hat\varphi(t,x))\!+\!\Delta \big(\frac{\sigma(t,x)^2}{2}\hat\varphi(t,x)\big),\!\!
\end{align}
\end{subequations}
with boundary conditions 
$$
\varphi(0,x)\hat\varphi(0,x)=p_0^{\rm exp}(x) \mbox{ and } \varphi(T,x)\hat\varphi(T,x)=p_T^{\rm exp}(x).
$$


This system of linear PDE's ({\em Schr\"odinger system}), where the Fokker-Planck equation for the prior dynamics~\eqref{eq:hatvarphi} and its adjoint \eqref{eq:varphi} are coupled through nonlinear boundary conditions, is known to have a unique solution under mild assumptions~\cite{fortet1940resolution,essid2019traversing}. Moreover, the solution can be obtained by a convergent algorithm due to Robert Fortet, which is also known as the Sinkhorn algorithm \cite{fortet1940resolution}, \cite[Section~8]{chen2021stochastic}. The algorithm consists in alternating between solving \eqref{eq:hatvarphi} forward in time and \eqref{eq:varphi} backward in time, using $p_0^{\rm exp}$ and $p_T^{\rm exp}$ to obtain the initial condition for one after computing terminal condition for the other. Schematically, this can be expressed as iterating the steps in the following diagram:
\begin{align}\label{eq:sinkhorn}\nonumber
\hat\varphi&(0,x)\quad\xrightarrow{\eqref{eq:hatvarphi}}\quad\hat\varphi(T,x)\\\tfrac{p_0^{\rm exp}(x)}{\varphi(0,x)}\ & \uparrow\qquad\qquad\qquad\qquad\downarrow\ \tfrac{p_T^{\rm exp}(x)}{\hat\varphi(T,x)}\\
\varphi&(0,x)\quad\xleftarrow{\eqref{eq:varphi}}\quad\varphi(T,x).\nonumber
\end{align}
Finally, having the probability density of $P^\star$ in the factored form \eqref{eq:factored}, leads to the optimal drift 
\begin{equation}\label{eq:control}
\beta(t,x)=\alpha(t,x)+\sigma(t,x)^2\nabla\log\varphi(t,x),
\end{equation}
corresponding to $P^\star$.


\subsection{Scope \& Significance of Schr\"odinger's problem}
The Schr\"odinger bridge can be interpreted as the solution to the following three problems that are equivalent but have a significantly different physical motivation.

\textbf{(i) A large deviations' problem.} This is along the lines of the Schr\"odinger's original {\em gedanken} experiment~\cite{Sch31,Sch32}. In this,
one seeks to account for unlikely events, that due to a finite number of diffusive particles, have given rise to endpoint marginals that are inconsistent with the law of large numbers dictating that solutions obey \eqref{eq:FP1}. Besides estimating the path traversed by the state of the system in this large deviations' scenario, Schr\"odinger was exploring the reversibility of physical laws and the nature of the solution in \eqref{eq:factored} that is reminiscent of the how probabilities are computed in quantum mechanics, as $\psi\bar\psi$, with $\psi$ a wave function.

\textbf{(ii) An inference problem.} This may be seen as a segue to the Maximum Caliber principle. Minimization of relative entropy between laws on paths can be thought of as a generalization of maximizing entropy to seek equilibrium distributions. Schr\"odinger's problem is analogous albeit in a dynamical context, where we seek to identify dynamics consistent with the observations at two endpoints.  Its solution allows us to infer the typical potential landscape (from \eqref{eq:control}) that gives rise to the most likely paths consistent with the endpoint measurements.

\textbf{(iii) A control problem.} Replacing $\beta(t,x)-\alpha(t,x)$
in \eqref{eq:dyn-SB} with $u(t,x)$, the integral 
       \begin{align}\nonumber
   & \int_0^T\int_{\mR^n}\frac{1}{\sigma^2(t,x)}\|u(t,x)\|^2 p(t,x)dxdt
\end{align}
can be interpreted as a penalty on high values of the control parameter $u$. Thereby, we may interpret Schr\"odinger's problem as the control problem that seeks the ``minimum energy'' control action $u$ that steers a stochastic ensemble of particles from a given initial to a given terminal distribution. 
In the simplest scenario when the prior is a Brownian diffusion ($\alpha(t,x)=0$) and the noise is homogeneous with $\sigma(t,x)=\epsilon$, the control problem takes the familiar form
      \begin{align}\nonumber
   & \min_{p,u} \int_0^T\int_{\mR^n}\|u(t,x)\|^2 p(t,x)dxdt
   \\[3pt]\nonumber
 & \quad \mbox{such that }p(0,\cdot)=p_0^{\rm exp},\ 
    p(T,\cdot)=p_T^{\rm exp}\\
    &\quad  \mbox{and } \,\partial_tp=-\nabla\cdot(up)+\frac12\Delta (\epsilon^2p).\label{eq:fpepsilon}
\end{align}
It's noteworthy that Schr\"odinger's problem can be viewed as an entropic regularization of the problem of Optimal Mass Transport with a quadratic cost \cite{leonard2012schrodinger,carlen2014stochastic,chen2016relation}, which can be written as in \eqref{eq:fpepsilon} for $\epsilon=0$. That is, Optimal Mass Transport is the zero-noise limit of Schr\"odinger's problem. The nonzero noise term renders the problem strictly convex, which facilitates computation, making the Schrodinger bridge problem extremely useful in solving problems of Optimal Mass Transport \cite{cuturi2013sinkhorn,chen2016entropic,chen2018efficient}.

As may have already become apparent, Schr\"odinger's bridge problem has emerged as a versatile tool in control theory and modeling of stochastic systems, but also in probability theory, data science, and machine learning, e.g. see \cite{chen2021controlling,chen2015optimal} and the references therein. In short, Schr\"odinger's bridge problem has proven key in theoretical and applied subjects where interpolation of distributions is of interest.

Paradigmatically, we mention Stochastic Thermodynamics \cite{sekimoto2010stochastic,bryan2022inferring,seifert2019stochastic}, where stochastic excitation models thermal fluctuation, and the Optimal Mass Transport problem with quadratic cost can be linked to entropy production \cite{aurell2011optimal}. Indeed, this framework allows us to obtain better estimates of the state of the ensemble, and therefore of thermodynamic currents at the level of the ensemble (such as work, heat, or entropy production), given endpoint information on the state of the system. Moreover, it provides a powerful tool to infer the potential landscapes that drive thermodynamic systems. Finally, in the noiseless limit, it allows finding trajectories that drive the system between two endpoints while minimizing entropy production.

This latter application area, Stochastic Thermodynamics, has motivated our study. Specifically, the experimental information that can be accounted for in the classical formulation of Schr\"odinger's problem is limited to distributions of the system at different points in time. Yet, typical experimental set-ups often allow continuous measurement of thermodynamic quantities such as energy, work, or heat \cite{ciliberto2017experiments}. In addition, one might only have access to information on certain degrees of freedom of the thermodynamic system (see e.g., the example in Section \ref{sec:ex1}), or on the population of different subsets of states (see the example in \ref{sec:ex2}). Motivated by such practical considerations, we extend Schr\"odinger's dictum to account for ensemble information of varying nature, and along paths of stochastic ensembles. In doing so, we bridge the theory of Schr\"odinger's problem to that of Maximum Caliber.

\section{A Schr\"odinger bridge approach to Maximum Caliber}\label{sec:MaxCalSB}

Schrödinger's bridge (SB) problem is intimately linked to the Maximum Caliber (MaxCal) formalism initiated by Jaynes half a century later \cite{jaynes1980minimum,presse2013principles}. 
Establishing a connection between these two problems provides an opportunity to leverage the conventional tools offered by classical Schr\"odinger bridges. Specifically, the framework we propose provides a dynamical description of optimal probability laws with the advantage of (i) being structured, (ii) being computationally approachable, and (iii) allowing us to infer and interpret the potential energy driving the system, under which the most likely observed dynamics are typical. 

Our development proceeds in two steps. 
First, we solve the maximum likelihood problem subject to a fixed path integral constraint and given initial and final boundary conditions ($p^{\rm exp}_0$ and $p^{\rm exp}_T$), just as in Schr\"odinger's problem. Then, we drop the boundary conditions to consider the standard MaxCal problem from a new perspective.
We conclude by specializing in cases where we seek to infer stationary distributions and corresponding potentials that are consistent with ensemble-path constraints.

\black

\subsection{Schr\"odinger bridges with currents}
\label{sec:BC}
In the spirit of Schr\"odinger's formalism we seek a law that is close to a prior in a relative entropy sense and agrees with observed marginal distributions, but this time, in addition, we introduce a constraint that represents integrated measurement along the duration of an experiment. We formulate a pertinent problem as follows:
\begin{subequations}\label{eq:SB-currents}
    \begin{align}
 & \min_{P:\,P\ll Q} \int_\Omega dP\log\frac{dP}{dQ}\\
 & \quad {\rm s.t.} \ \int_\Omega dP=1,\   p_0=p_0^{\rm exp},\ p_T=p_T^{\rm exp}\\
& \quad \mbox{and } \int_\Omega \int_0^T j(t,X_t) dt dP=\text{\j}^{\rm exp}.\label{eq:current-t}
\end{align}
\end{subequations}
Here, $\int_0^Tj(t,X_t)dt\in\mathbb R$ represents a current being measured 
over the duration of the experiment and \eqref{eq:current-t} its average at
the level of the ensemble. Such an integral
may represent energy/work \cite[Chapter 4]{sekimoto2010stochastic} or other moments, possibly measured along specific degrees of freedom, or population (or probability) in a certain subset of $\mathbb R$.
For simplicity of notation and without loss of generality we consider one such integral.
A dynamic constraint where the ensemble average
\begin{equation}\tag{\ref{eq:SB-currents}c'}
\label{eq:dyn-constr}
    \int_\Omega  j(t,X_t)dP=\text{\j}^{\rm exp}(t)
\end{equation}
is specified across time, can also be treated quite similarly.
Such an average may model the time history of the work rate, for example.
 We will point out whenever an argument needs to be modified to treat dynamical constraints in the form of \eqref{eq:dyn-constr}.

\subsubsection{Laws on paths}\label{sec:BC-paths}
From a path-space point of view, the Lagrangian $\mathcal L$ for this optimization problem can be written as
$$
\int_\Omega\! \Big(\log\frac{dP}{dQ}+\gamma+\mu(X_0)+ \nu(X_T)+\!\int_0^T \!\!\lambda j(t,X_t) dt \Big)dP.
$$
Its first variation $\delta\mathcal L(\Lambda;\delta)$ with respect to $\Lambda=\frac{dP}{dQ}$, i.e. $\mathcal L(\Lambda+\delta)-\mathcal L(\Lambda)$, is
$$
\int_\Omega \!\!\Big(\log\Lambda +\gamma+1+\mu(X_0)+ \nu(X_T)+\!\int_0^T \!\!\lambda j(t,X_t) dt \Big)\delta dQ.
$$
Thus, the first order stationarity condition where the variation vanishes for all $\delta$, gives that the optimal choice for $\Lambda$ is
$$
    \Lambda^\star=\frac{e^{-\mu(X_0)}e^{-\int_0^T \lambda j(t,X_t) dt}e^{- \nu(X_T)}}{\mathcal Z},
$$
where $\mathcal Z=e^{\gamma+1}$ is a normalization constant (partition function). Hence,
\begin{equation}\label{eq:fegQ}
P^\star=f(X_0)e^{-\int_0^T \lambda j(t,X_t) dt}g(X_T) Q,
\end{equation}
where $f(X_0)\propto e^{-\mu(X_0)}$ and $g(X_T)\propto e^{-\nu(X_T)}$, having absorbed the normalization constant into $f,g$. 

Equation \eqref{eq:fegQ} displays a form of Doob's h-transform \cite{doob1957conditional} (also Feynnman-Kac \cite{leonard2022feynman}). Such transforms are central to problems of large deviations and preserve the Markovian character of $Q$~\cite{leonard2011htransf,chetrite2015variational}. 
Variants have also been utilized in Schr\"odinger bridge problems where endpoint marginals have different mass (and, therefore, creation or killing of particles has taken place) \cite{chen2022most}. 

We finally note that a similar analysis applies to the case where the dynamical constraint \eqref{eq:dyn-constr} is imposed, instead of \eqref{eq:current-t}.
The only difference between the two is that in the case of  \eqref{eq:dyn-constr},
the Lagrange multiplier $\lambda$ needs to be a function of $t$, as opposed to being a scalar.

\subsubsection{A stochastic control perspective}\label{sec:BC-SC}
To gain an intuition for the nature of \eqref{eq:fegQ}, let us consider the corresponding stochastic dynamics and process.
Starting from the diffusion process \eqref{eq:p-dyn} with law $Q$, the process corresponding to the law  \eqref{eq:fegQ} is Markovian and obeys
\begin{equation}\label{eq:forward-x}
    dX_t=\beta(t,X_t)dt+\sigma(t,X_t)dW_t.
\end{equation}
Applying once again Girsanov's theorem, \eqref{eq:SB-currents} can be recast as the following minimization problem:
\begin{subequations}
\begin{align}\label{eq:dyn-SB-c}
   & \min_{p,\beta} \frac12\int_0^T\int_{\mR^n}\frac{\|\beta(t,x)-\alpha(t,x)\|^2}{\sigma(t,x)^2}p(t,x)dxdt,
\end{align}
subject to
    \begin{gather}\label{eq:current}
 \int_0^T\int_{\mR^n} j(t,x)p(t,x)dxdt=\text{\j}^{\rm exp},\\\label{eq:fp}
\partial_tp=-\nabla\cdot(\beta p)+\frac12\Delta (\sigma^2p)\\\label{eq:bridge-BC}
    \mbox{and } \,      p(0,\cdot)=p_0^{\rm exp},\ 
    p(T,\cdot)=p_T^{\rm exp}.
\end{gather}
\end{subequations}

This time the Lagrangian $\mathcal L$ is
\begin{align}\label{eq:lagrangian}
&\int_0^T\!\int_{\mR^n}  \Bigg\{\frac12\frac{\|\beta-\alpha\|^2}{\sigma^2}p+\lambda j p\Bigg\} dxdt-\text{\j}^{\rm exp}\\&+\int_0^T\!\int_{\mR^n} \Bigg\{\log\varphi\Big( \partial_tp+\nabla\cdot(\beta p)-\frac12\Delta( \sigma^2p)\Big)
\Bigg\} dxdt.\nonumber
\end{align}
Here, $\log\varphi(t,x)$ is a Lagrange multiplier introduced to ensure that the Fokker-Planck equation \eqref{eq:fp} holds.

The vanishing of the first variation with respect to $\beta$ is both a necessary as well as sufficient condition for optimality, due to strict convexity.
Specifically, the first variation with respect to $\beta$ reads
\black
\begin{align}\label{eq:delta-b}
   &\delta\mathcal L(\beta;\delta_\beta)= 
   \int_0^T\int_{\mR^n}\Big(\frac{\beta-\alpha}{\sigma^2}-\nabla\log\varphi\Big)'p\delta_\beta dxdt.
\end{align}
Setting this to zero for all functions $\delta_\beta$, we obtain
\begin{align}\label{eq:beta}
\beta(t,x)&=\alpha(t,x)+\sigma(t,x)^2\nabla\log\varphi(t,x).
\end{align}
We see that the posterior drift differs from the prior by the term $\sigma^2\nabla\log\varphi$. 
This has exactly the same structure as in~\eqref{eq:control}. 
In the important case where  $\alpha$ is a gradient of a potential, and provided $\sigma$ is scalar and independent of $x$, the posterior drift $\beta$ is also the gradient of a potential.
The difference between \eqref{eq:beta} and \eqref{eq:control} comes in through the $\varphi$ dynamics.
Indeed, the first variation with respect to $p$~is
\begin{align*}
   \delta \mathcal L(p,\delta_p)= 
   \int_0^T\!\!\int_{\mR^n}
\Big(&-\partial_t\log\varphi-\beta'\nabla\log\varphi-\frac{1}{2}\sigma^2\Delta\log\varphi\\&+\frac12\|\sigma\nabla\log\varphi\|^2+\lambda j\Big)\delta_p
     dxdt,
\end{align*}
 where we have used \eqref{eq:beta} and integration by parts. 
Setting the integrand to zero, 
 \begin{align}
\partial_t\log\varphi\!&=\nonumber\!
    -\alpha'\nabla\log\varphi-\frac12\|\sigma\nabla\log\varphi \|^2\!-\frac12\sigma^2\Delta\log\varphi+\lambda j\\&=-\alpha'\nabla\log\varphi-\frac12\sigma^2\frac{\Delta\varphi}{\varphi}+\lambda j,\label{eq:logphi}
\end{align}
where for the second equality we have used the identity $\Delta\log \varphi+\|\nabla \log \varphi\|^2=\Delta \varphi/\varphi$, we obtain
\begin{subequations}\label{eq:Schrsystem}
\begin{equation}\label{eq:phi}
    \partial_t\varphi=-\alpha'\nabla\varphi-\frac12\sigma^2\Delta\varphi+\lambda j\varphi.
\end{equation}

The solution for \eqref{eq:phi} dictates our new drift $\beta$. However, we still need to find the appropriate terminal condition $\varphi(T,x)$ that will allow matching the boundary conditions \eqref{eq:bridge-BC}.
To do so, it is convenient to define
$$
\hat\varphi(t,x)=\frac{p(t,x)}{\varphi(t,x)}.
$$ 
Taking the time derivative,
$
\partial_t\hat\varphi=\partial_tp/\varphi-\hat\varphi\partial_t\varphi/\varphi,
$
and using the dynamical equations for $p$ and $\varphi$, \eqref{eq:fp}  and \eqref{eq:phi} respectively, we obtain
\begin{align}\label{eq:hat-dyn}
\partial_t\hat\varphi&=-\nabla\cdot(\alpha \hat\varphi)+\frac12\Delta (\sigma^2\hat\varphi)-\lambda j\hat\varphi.
\end{align}
\end{subequations}
Therefore, instead of solving \eqref{eq:fp}
and \eqref{eq:phi}, we can solve the Schr\"odinger system (\ref{eq:phi}-\ref{eq:hat-dyn}), together with boundary conditions
$\varphi(T,x)$ and $\hat\varphi(0,x)$ such that 
$$
\varphi(0,x)\hat\varphi(0,x)=p_0^{\rm exp}(x) \mbox{ and } \varphi(T,x)\hat\varphi(T,x)=p_T^{\rm exp}(x).
$$

In \cite{chen2022most}, it is shown that for each choice of $p_0^{\rm exp},\,p_T^{\rm exp}, \,\lambda$ (and for enough regularity in $j(t,x)$) there exists a unique pair of $\hat\varphi(0,x),\, \varphi(T,x)$ (that evolve according to (\ref{eq:phi}-\ref{eq:hat-dyn})), such that 
\begin{equation}\label{eq:p-sol}
    p(t,x)=\varphi(t,x)\hat\varphi(t,x)
\end{equation}
 satisfies the boundary conditions $p_0^{\rm exp}$ and $p_T^{\rm exp}$. This pair can be found via an iterative algorithm akin to~\eqref{eq:sinkhorn}, in which the only difference is the term proportional to $\lambda j(t,x)$ in \eqref{eq:phi} and \eqref{eq:hat-dyn}.
In \cite{chen2022most}, $\lambda j(t,x)$ represented the killing rate of particles in an unbalanced Schr\"odinger bridge problem in which particles can appear and disappear. In this work, the ``creation'' and ``annihilation'' terms 
in (\ref{eq:phi},\ref{eq:hat-dyn}) cancel 
out to obtain \eqref{eq:p-sol}, that conserves mass. In addition, 
$\lambda$ constitutes a Lagrange multiplier that enforces the ensemble constraint~\eqref{eq:current}. Therefore, we are also required to iterate to find the optimal $\lambda$ that achieves the required current. 

Before we do that, note that for each fixed $\lambda$ the obtained solution is indeed in the form of \eqref{eq:fegQ}. To see this, let us write the Radon-Nikodym derivative between $P$ and $Q$ with the optimal drift $\beta$ \eqref{eq:beta} as
\begin{equation}
    \label{eq:RN1}
    \frac{dP}{dQ}=\frac{p^{\rm exp}_0}{q_0}e^{\frac12\int_0^T\|\sigma\nabla\log\varphi\|^2dt +\int_0^T \sigma\nabla\log\varphi'{dW_t}}.
\end{equation}
Using the Itô rule to take the differential of $\log\varphi(t,X_t)$ we realize that
\begin{align}\label{eq:ito}
    \sigma\nabla\log\varphi' dW_t=&-\partial_t\log\varphi dt -\beta'\nabla\log\varphi dt-\frac12\sigma^2\Delta\log\varphi \nonumber\\&+d\log\varphi.
\end{align}
Substituting $\partial_t\log\varphi$ by \eqref{eq:logphi}, we obtain that \eqref{eq:RN1} can be written as
\begin{align*}
    \frac{dP}{dQ}&=\frac{p^{\rm exp}_0(X_0)}{q_0(X_0)}e^{\log\varphi(T,X_T)-\log\varphi(0,X_0)-\int_0^T\lambda j(t,X_t) dt}\\&
    =\underbrace{\frac{\hat\varphi(0,X_0)}{q_0(X_0)}}_{f(X_0)}e^{- \int_0^T\lambda j(t,X_t) dt}\underbrace{\varphi(T,X_T)}_{g(X_T)},
\end{align*}
matching \eqref{eq:fegQ}.

\subsubsection{Computing the Lagrange multiplier}\label{sec:lagrange}
It remains to determine the Lagrange multiplier $\lambda$ that enforces the constraint
\begin{equation}\label{eq:constr-h}
    h(\lambda):=\int_\Omega \int_0^T j(t,X_t) dtdP=\text{\j}^{\rm exp}.
\end{equation}
Note that, using \eqref{eq:fegQ},
$$
\partial_\lambda h(\lambda)=- \int_\Omega\bigg(\int_0^T j(t,X_t) dt\bigg)^2dP\leq 0.
$$
 Thus, $h$ is a monotone function of $\lambda$, as expected from the convexity of the problem. 
 Therefore, we can have a simple iterative way of finding $\lambda$, such as  Newton's method or gradient descent in the dual $\lambda$-space, in which, at each iteration, equations \eqref{eq:phi} and \eqref{eq:hat-dyn} need to be solved for a new value for $\lambda$. The non-existence of an optimal $\lambda$ is not of concern here, since it is assumed that any physically meaningful measurement will lead to an attainable ensemble constraint.

 Note that, if instead of imposing a time-integrated constraint like \eqref{eq:current-t}, we are interested in a rate-constraint as in \eqref{eq:dyn-constr}, $\lambda$ needs to be a function of time. The search for $\lambda(t)$ is then more involved since it becomes an infinite dimensional search. Nevertheless, it is still approachable through gradient descent in the dual space. Specifically, we seek $\lambda(t)$ as the extremum of \eqref{eq:lagrangian}. That is, in each iteration we set
\begin{align}\label{eq:grad-asc}
\lambda(t)\to\lambda(t)+s\!\times\!\bigg(\int_{\mathbb R^n}j(t,x)p(t,x)dx- \text{\j}^{\rm exp}(t)\bigg),
\end{align}
where $s$ is the step size of the gradient descent.



\subsubsection{Scope \& Significance}

We have derived a dynamical description \eqref{eq:p-sol} of the most likely path that satisfies two boundary conditions and an ensemble-path constraint. It allows updating a prior model to one that is consistent with measurements and thereby provides improved estimates of other physical quantities of interest. 
At the same time, it provides an update of the drift term \eqref{eq:beta} that, when the diffusion coefficient is independent of $x$, is in gradient form. Hence, an updated potential landscape that drives the system dynamics is revealed, giving a better understanding of the underlying  properties of the physical, chemical, or biological system.

The problem at hand can also be interpreted as one seeking a control action $u=\beta-\alpha$ with minimum energy, that not only drives a stochastic system between two specified endpoint distributions but also ensures ensemble constraints along the path. Its noise-less limit ($\sigma\to 0$)
    \begin{align}\nonumber
   & \min_{p,u} \int_0^T\int_{\mR^n}\|u(t,x)\|^2 p(t,x)dxdt\\[3pt]\nonumber
 & {\rm s.t.} \   p(0,\cdot)=p_0^{\rm exp},\ 
    p(T,\cdot)=p_T^{\rm exp},\\\nonumber
    &\quad \ \ \int_0^T\int_{\mR^n}j(t,x)p(t,x)dxdt=\text{\j}^{\rm exp},\\\nonumber
    &\quad \ \ \mbox{and } \,\partial_tp=-\nabla\cdot(up), 
\end{align}
represents a variant of the standard Optimal Mass Transport (OMT) problem in which integral path constraints are introduced.
Such problems are particularly relevant in Stochastic Thermodynamics, where $u(t,x)$ typically represents the mean velocity of particles and the quadratic cost on $u$ represents entropy production along the path from $p_0^{\rm exp}$ to $p_T^{\rm exp}$ \cite{aurell2011optimal,chen2019stochastic}.
Therefore, this OMT problem is simply that of minimizing entropy production while driving the system between two endpoints with a fixed current $\text{\j}^{\rm exp}$, which could represent, for example, work produced along a trajectory. 

There have been previous efforts to unify OMT theory and MaxCal in the discrete-time and discrete-space setting \cite{dixit2019building}.
Therein, the entropic regularization term is used to account for entropy minimization, while the original OMT cost represents an ensemble average constraint, analogous to \eqref{eq:dyn-constr}. Thus, the standard discrete OMT problem is used to solve a certain MaxCal problem, without additional constraints. In this work, they derive optimal transition rates by assuming knowledge of initial and final distributions \textit{at every time step}. 
However, such an approach does not appear suitable in our setting.
Herein, ensemble-path constraints are added to the entropy-regularized OMT problem. Moreover, the optimization is over paths in the probability space; these are not given, only their endpoints are fixed.

\subsection{The Maximum Caliber problem}

In typical experimental setups, one may not have access to complete knowledge of the distribution at time-end points.
Thus, we now consider the more standard MaxCal problem in which we minimize relative entropy subject to observed currents, but otherwise with no additional constraints. That is, we consider the problem
\begin{align}
   & \min_{P:\,P\ll Q} \int_\Omega dP\log\frac{dP}{dQ}\\\nonumber
 &\ \  {\rm s.t.} \ \int_\Omega \int_0^T j(t,X_t) dt dP=\text{\j}^{\rm exp}\mbox{ and } \int_\Omega dP=1.
\end{align}
Note that in standard Maximum Caliber problems, entropy over paths is minimized (instead of relative entropy). This simply amounts to the prior $Q$ being    ``uniform'' over $\Omega=\mathcal C([0,T])$, e.g., the stationary Wiener measure \cite{leonard2014some}.
Then, in the spirit of Maximum Entropy, the problem corresponds to finding the optimal $P$ that would be uniform over all possible trajectories (the dynamical equivalent of microstates) were it not for the fact that a current (the dynamical equivalent of energy) is fixed. 

Following the same steps as in Section~\ref{sec:BC-paths}, it is clear that the optimal law on paths $P$ must be given by
\begin{equation}\label{eq:RN2}
   \frac{dP^\star}{dQ}=\frac{e^{-\int_0^T \lambda j(t,X_t) dt}}{\mathcal Z},  
\end{equation}
where $\mathcal Z$ is the normalization constant. This structure is typical of Maximum-Caliber problems and is such that $P^\star$ has a Markovian structure as long as $Q$ does too.  

Consider now a diffusion process \eqref{eq:p-dyn} with drift $\alpha$ as our prior. Due to Markovianity, the posterior will be the diffusion process in \eqref{eq:forward-x} with drift $\beta$. Instead of going the same route as in the previous section, we will directly find the stochastic process that corresponds to the law defined by \eqref{eq:RN2} \cite{leonard2022feynman}. In this case, this turns out to be considerably easier than solving the optimization problem from the dynamic point of view (as in Section \ref{sec:BC-SC}), since the boundary conditions on $\varphi$ and $\hat\varphi$ will come in naturally, as we will see next.

Let us postulate that the optimal drift $\beta$ is again be of the form $\alpha+\sigma^2\nabla\log\varphi$, with $\varphi$ evolving according to~\eqref{eq:phi}.
The Radon-Nikodym derivative between $P$ and $Q$ with this $\beta$ reads
\begin{equation}
    \frac{dP^\star}{dQ}=\frac{p_0}{q_0}e^{\frac12\int_0^T\|\sigma\nabla\log\varphi\|^2dt +\int_0^T \sigma\nabla\log\varphi'{dW_t}}.
\end{equation}
Using \eqref{eq:ito}, and substituting $\partial_t\log\varphi$ by \eqref{eq:logphi}, we obtain, as before,
\begin{align}\label{eq:RN3}
    \frac{dP^\star}{dQ}&=\frac{\hat\varphi(0,X_0)}{q_0(X_0)}\varphi(T,X_T)e^{- \int_0^T\lambda j(t,X_t) dt},
\end{align}
where $\hat\varphi(0,X_0)=\frac{p_0(X_0)}{\varphi(0,X_0)}$.
However, this expression must match \eqref{eq:RN2}. This implies that, up to a multiplicative constant,
$$
\hat\varphi(0,X_0)=q_0(X_0)\ \mbox{ and }\ \varphi(T,X_T)=\frac{1}{\mathcal Z},
$$
since the expression on the right-hand side of \eqref{eq:RN3} can not be a function $X_0$ or $X_T$.
Therefore, the optimal one-time marginals are given by
$$
p(t,x)=\varphi(t,x)\hat\varphi(t,x),
$$
with $\varphi(t,x),\,\hat\varphi(t,x)$ satisfying
\begin{subequations}
    \begin{align}\label{eq:MCphi}
           \partial_t\varphi&=-\alpha'\nabla\varphi-\frac12\sigma^2\Delta\varphi+\lambda j\varphi,\ \varphi(T,\cdot)=\frac{1}{\mathcal Z}\\
           \partial_t\hat\varphi&=-\nabla\cdot(\alpha \hat\varphi)\!+\!\frac12\Delta (\sigma^2\hat\varphi)-\lambda j\hat\varphi,\ \hat\varphi(0,\cdot)=q_0,
    \end{align}
    where $q_0$ is the one-time marginal of $Q$ at $t=0$, as before.
\end{subequations}

Note that, in contrast to the Schr\"odinger bridge problem, here the boundary conditions are uncoupled, and thus there is no need for (Sinkhorn-type) iteration. In practice, however, $\mathcal Z$ is unknown. Thus, one can solve \eqref{eq:MCphi} with $\varphi(T,\cdot)=1$ as the boundary condition.
Then, the normalization constant $\mathcal Z$ can be simply found at any time as $\mathcal Z=\int \varphi(t,x)\hat\varphi(t,x)dx$, leading to the optimal solution $p(t,x)=\varphi(t,x)\hat\varphi(t,x)/\mathcal Z$.
On the other hand, the Lagrange multiplier $\lambda$ remains to be found, which must be done iteratively as in Section \ref{sec:lagrange}. Once again, it is possible to account for multiple current constraints (leading to vectorial $\lambda$),  as well as dynamic constraints of the form of \eqref{eq:dyn-constr} (leading to time-varying $\lambda(t)$). 

To sum up, we have found an explicit dynamical description of the most likely paths and the typical potential landscapes that give rise to the observed current. This has the advantage of being both physically interpretable and computationally approachable. Moreover, this solution still solves an optimal control problem in which control effort is minimized. However, this control effort is no longer used to drive the system between endpoint states, but only to attain a certain  ensemble-path average (e.g. a moment of the distribution), regardless of initial and final distributions.

A recent work in this direction \cite{bolhuis2022force} has aimed to obtain the most likely constant parametric perturbation of a prior potential energy given a dynamical constraint, specifically, a rate constant. Therein, implicit necessary conditions for the optimal potential parameters are provided, where up to two parameters are considered. On the other hand, our framework is general in terms of the possible posterior potentials, which are allowed to be time-varying. This provides a higher level of abstraction that allows for a clear structure of the optimal solutions. Other works have also inferred the transition probabilities from steady-state data \cite{dixit2014inferring,dixit2015inferring,dixit2018caliber}, and postulate that these transition probabilities still hold out of equilibrium. Herein, however, we update our model with possibly non-stationary data and allow for time-dependent drifts (analogous to time-dependent transition probabilities, as we will see in Section \ref{sec:discrete}) that better match non-stationary dynamics. Thus, we provide a novel framework for dynamical Maximum Caliber-type model inference, in which kinetic/thermodynamic macroscopic properties
give force fields reporting on the underlying
physics.

\subsection{Steady-state}
The introduced framework can also be used to infer stationary dynamics that are consistent with ensemble constraints. Specifically, the steady-state arises as a particular case by assuming that $\partial_tp=0$ and no boundary conditions, that is, no information on the steady-state $p_{\rm ss}$ of the system. The prior and the observed current must also be stationary, in the sense that we now assume $\alpha(t,x)=\alpha(x),$ $\sigma(t,x)=\sigma(x)$ and 
\begin{equation}\label{eq:constr-ss}
\int_{\mR^n}j(x)p_{\rm ss}(x)dx=\text{\j}^{\rm exp}.
\end{equation}
By following the same steps as in Section \ref{sec:BC-SC}, we obtain the optimal drift
$$
\beta=\alpha+\sigma^2\nabla\log\varphi,
$$
together with the steady-state Schr\"odinger system
\begin{subequations}\label{eq:ss-cont}
    \begin{align}
        \alpha\nabla\varphi+\frac12\sigma^2\Delta\varphi-\lambda j \varphi=0,\\
-\nabla\cdot(\alpha \hat\varphi)+\frac12\Delta (\sigma^2\hat\varphi)-\lambda j\hat\varphi=0.
\end{align}
\end{subequations}
Thus, $\varphi$ and $\hat\varphi$ are the zero eigenfunctions of a ``weighted'' infinitesimal generator  of $Q$, $\alpha \nabla +\frac12\sigma^2\Delta - \lambda j$, and its adjoint, respectively.
\black
Assuming enough regularity of $\alpha$ and $j$, these equations can be solved for $\varphi$ and $\hat\varphi$, to give the most likely steady-state of the system
$$
p_{\rm ss}(x)=\varphi(x)\hat\varphi(x),
$$
which indeed satisfies 
$$
\partial_tp_{\rm ss}=0=-\nabla\cdot(\beta p_{\rm ss} )+\frac12\Delta(\sigma^2p_{\rm ss}).
$$
Previous works have obtained analogous expressions for the discrete case \cite{dixit2015stationary,dixit2018communication,PhysRevE2019ss} (see Section \ref{sec:discrete-ss}), whereas in this work these appear as a special case of the general dynamic framework.
Note that, when $\sigma$ is a constant scalar, as long as the prior satisfies detailed balance, the posterior does too, since the only update is a gradient term on the drift \footnote{Here, by detailed balance we understand the vanishing of the probability current. For the $\sigma$ constant case, since $\partial_t p_{\rm ss}=-\nabla\cdot J=0$, the probability current $J$
$$
J:=\big(\alpha+\sigma^2\nabla\log\varphi-\frac12\sigma^2\nabla\log p_{\rm ss}\big)p_{\rm ss}=0,
$$
whenever the prior is detailed balanced, this is, whenever $\alpha$ is of gradient form}.
 This may no longer be true in general when $\sigma$ is a matrix or a function of $x$.


It is important to note that the problem tackled in this subsection is not equivalent to a naive maximum entropy approach in which one may seek the steady-state distribution that is closest to the prior in relative entropy while satisfying certain ensemble-average constraint. Namely,
$$
{\rm arg}\min_{p_{\rm ss}}\int_{\mR^n}p_{\rm ss}(x)\log\frac{p_{\rm ss}(x)}{q_{\rm ss}(x)}dx,\ \mbox{subject to \eqref{eq:constr-ss}.}
$$
Such an approach only outputs a steady-state distribution $p_{\rm ss}$ which is consistent with an infinite number of underlying dynamics. Then, in a separate optimization problem, one would need to find the most likely dynamics leading to this steady state, resulting in an overall sub-optimal solution. Instead, herein,
the steady-state distribution {\em and} the dynamics are simultaneously optimized.


\section{Discrete Setting}
\label{sec:discrete}

So far we have developed our theory
in continuous time and continuous state space.
In the present section, we follow up with the counterpart for dynamics taking place on a discrete space and over a discrete time-indexing set. Relevant applications abound, as in   
chemical networks~\cite{ghosh2020maximum}, genetic circuits~\cite{firman2017building}, systems that model the firing of neurons~\cite{neurons2015maxcal} or conformational changes of molecules and biopolymers~\cite{wan2016maximum}, and many others.
However, an added incentive for studying the discrete setting is drawn from parallels to the continuous counterpart of the theory that ultimately helps elucidate both.
Hence, we now consider the discrete time/space  Schr\"odinger-Maximum Caliber problem.

Let $\mathcal X$ denote a discrete state space that, for simplicity, we assume finite, i.e., with cardinality $|\mathcal X|=n$. Then, $\mathcal X^{T+1}$ is the space of sample paths with $T+1$ time steps, with the associated sample points denoted as
\begin{align*}
    \mathbf x = (x_0, x_1, \cdots, x_T), 
\end{align*}
for $x_0, x_1, \cdots, x_T \in \mathcal X$. In other words, a path $\mathbf x$ is a function from the time indexing set $\{0,1,\ldots,T\}$ to $\mathcal X$.

Let $Q$ be a (prior) Markov Law on $\mathcal X^{T+1}$, specified by
\begin{align*}
    Q( \mathbf x) = q_0(x_0) q_{x_1 \mid x_0}  q_{x_2 \mid x_1}  \cdots q_{x_T \mid x_{T-1}},
\end{align*}
where $q_0$ is an initial probability distribution (vector) on $\mathcal X$ and $q_{x_t \mid x_{t-1}}$ represents the transition probability to a state $x_t$ from a state $x_{t-1}$ at time $t$. 
The Markov transition kernel satisfies 
\begin{align*}
    \sum_{x_t \in \mathcal X} q_{x_t \mid x_{t-1}} =1,
\end{align*}
for $t \in \{0,1,\cdots, T\}$.
%
As compared to previous related works~\cite{dixit2014inferring,dixit2015inferring}, with little overhead, we develop our framework assuming time-varying prior, for generality. 

As before, we assume that our prior $Q$ is inconsistent with measurements (ensemble-path integrals in the form of currents), and our task is to determine the most likely law that is compatible with the measurements.
The optimal posterior law is foreseeably Markov, similar to the continuous case. Indeed, as shown next, the discrete optimization problem yields a new Markov kernel that leads to the most likely sought stochastic evolution.

\subsection{Schr\"odinger bridges with currents}
We first consider the Schr\"odinger bridge problem with a current constraint which, in analogy to the continuous case, takes the form
\begin{align}\label{eq:discrete-SBP}
   & \min_{P \ll Q}  \sum_{\mathbf x \in \mathcal X^{T+1}} P (\mathbf x) \log \frac{P (\mathbf x)}{Q (\mathbf x)}, \\[2pt]\nonumber
   & {\rm s.t.}  \ p_0(x_0) = p_0^{\rm exp}(x_0),    \ p_T(x_T)= p_T^{\rm exp}(x_T),\\[2pt]
   &   \sum_{x_0, \cdots, x_T \in \mathcal X} \left(\sum_{k=0}^{T-1} j_k(x_k, x_{k+1}) \right)P(x_0, x_1 ,\cdots, x_T) = \text{\j}^{\rm exp}.\nonumber
\end{align}
Here, $j_k(x_k, x_{k+1})$ denotes the measured current between sites $x_k$ and $x_{k+1}$ in $\mathcal X$, and $p_t$ denotes the one-time marginal for the distribution $P$, that is,
\begin{align*}
   p_t(x_t) =  \sum_{x_1, \cdots, x_T \setminus x_t} P(x_0, x_1 ,\cdots, x_T).
\end{align*}
The measured current may capture quantities of interest such as work, heat, dwell times, or traffic between two states \cite{peliti2021stochastic}. 


Based on similar arguments presented in the continuous setting, it is straightforward to verify that the optimizer takes the form 
\begin{equation}\label{eq: optdis}
     P^\star (\mathbf x) =  f(x_0)g(x_T)  e^{- \sum\limits_{k=0}^{T-1} \lambda j_k(x_k, x_{k+1})} Q(\mathbf x),
\end{equation}
where $f(x_0)$ and $g(x_T)$ arise as Lagrange multipliers of the endpoint constraints $p_0^{\rm exp}$ and $p_T^{\rm exp}$.  The structure of $P^\star$ implies that the Markovianity of $Q$ is preserved.
This can be traced to the fact that the Radon-Nikodym derivative factors into a product of $2$-tensors, since the exponent is likewise the sum of $2$-tensors.
\black
Imposing the endpoint constraints we obtain
\begin{subequations}\label{eq:constr-phi}
    \begin{align}\nonumber
\hat\varphi_0(x_0)\!\!\!\!\sum_{x_1,\cdots,x_{T}}\!\!\!\!\varphi_T(x_T)  &e^{- \sum\limits_{k=0}^{T-1} \lambda j_k(x_k, x_{k+1})} q_{x_1 \mid x_0} \cdots q_{x_T \mid x_{T-1}}\\&\qquad\qquad\qquad\ \ \ \  =p_0^{\rm exp}(x_0)\\\nonumber
     \varphi_T(x_T)\!\! \!\!\!\!\sum_{x_0,\cdots,x_{T-1}} \!\!\!\!\!\! \!\hat\varphi_0(x_0) &e^{- \sum\limits_{k=0}^{T-1} \lambda j_k(x_k, x_{k+1})} q_{x_1 \mid x_0} \cdots q_{x_T \mid x_{T-1}}\\&\qquad\qquad\qquad\ \ \ \  =p_T^{\rm exp}(x_T),
\end{align}
\end{subequations}
where we have defined
$$
\hat\varphi_0(x_0):=f(x_0) q_0(x_0),\ \mbox{ and } \ \varphi_T(x_T) :=g(x_T).
$$
The structure of the equations suggests an evolution
\begin{subequations}\label{eq:dis-Ssys}
    \begin{align}
&\hat \varphi_{t+1}(x_{t+1}) := \sum_{x_t}  q_{x_{t+1} \mid x_t} \Lambda(x_t,x_{t+1}) \hat \varphi_t(x_t) \label{eq: fordis}, \\
 &   \varphi_t(x_t) := \sum_{x_{t+1}}  q_{x_{t+1} \mid x_{t}} \Lambda(x_t,x_{t+1}) \varphi_{t+1}(x_{t+1}) \label{eq: backdis},  
\end{align}
 for $\hat\varphi$ and $\varphi$, with $t\in\{0,\cdots,T\}$ and $\Lambda(x_k,x_{k+1}) := e^{- \lambda j_k(x_k,x_{k+1})}$.  Then, equations \eqref{eq:constr-phi} couple the boundary conditions of \eqref{eq:dis-Ssys}, since
   \begin{align}
   \hat \varphi_0 (x_0)  \varphi_0(x_0)&= p_0^{\rm exp}(x_0), \label{eq: const1}\\
    \varphi_T(x_T) \hat \varphi_T(x_T) &= p_T^{\rm exp}(x_T) \label{eq: const2}. 
\end{align}
\end{subequations}
The system \eqref{eq:dis-Ssys} is the discrete Schr\"odinger system \cite{pavon2010discrete,georgiou2015positive}.
For $\lambda$ specified,
the uniqueness of solutions follows by the same argument as in \cite{georgiou2015positive}\footnote{One simply has to notice that 
the map
\begin{align*}
   \hat  \varphi_0^i \stackrel{\eqref{eq: fordis}}\mapsto \hat \varphi_T^i \stackrel{\eqref{eq: const2}}\mapsto \varphi_T^i \stackrel{\eqref{eq: backdis}}\mapsto \varphi_0^i \stackrel{\eqref{eq: const1}}\mapsto \hat  \varphi_0^{i+1} 
\end{align*}
for $i=1, 2, \ldots$ is a contraction in the Hilbert metric when $q_{x_t \mid x_{t-1}}$ is positive for $t=0, \ldots, T$.}.
We note that equation \eqref{eq: fordis} and \eqref{eq: backdis} are the discrete evolution counterparts of the Fokker-Planck equation \eqref{eq:hat-dyn} and its adjoint \eqref{eq:phi}, respectively.
A point of departure from these earlier works \cite{pavon2010discrete,georgiou2015positive}  is that a parameter $\lambda$ is to be determined so as to satisfy the last ensemble path condition in \eqref{eq:discrete-SBP}.

Summing over all $x_i\in\mathcal X$ but for $x_t$ in \eqref{eq: optdis}, the one-time marginals of $P^\star$ can be computed to be
\begin{align}\label{eq:dis-phiphihat}
    p_t(x_t) = \varphi_t(x_t) \hat \varphi_t(x_t),
\end{align}
for all $t \in \{0, 1, \cdots, T\}$.
Moreover, the optimal law can be factored into
\begin{align}
    P^\star(\mathbf x):= p_0(x_0) p_{x_1 \mid x_0}  p_{x_2 \mid x_1}  \cdots p_{x_T \mid x_{T-1}},
\end{align}
by noticing
\begin{align*}
\begin{split}
 P^\star(\mathbf x) =& \hat \varphi_0(x_0) q_{x_1 \mid x_0} \Lambda(x_0,x_1) \times  \cdots   \\
     & \times q_{x_T \mid x_{T-1}} \Lambda(x_{T-1},x_T) \varphi_T(x_T),\\
=&  \varphi_0(x_0) \hat \varphi_0(x_0) q_{x_1 \mid x_0} \Lambda(x_0,x_1) \frac{\varphi_1(x_1)}{\varphi_0(x_0)}  \\
    & \times  \frac{1}{\varphi_1(x_1)} \times  \cdots   \times \varphi_{T-1}(x_{T-1})  \\
    &\times q_{x_T \mid x_{T-1}} \Lambda(x_{T-1},x_T) \frac{\varphi_T(x_T)}{\varphi_{T-1}(x_{T-1})}.
    \end{split}
\end{align*}
Therefore, the new Markov kernels are given by
\begin{align}\label{eq:markovdis}
    p_{x_{t+1} \mid x_t} = q _{x_{t+1} \mid x_t} \Lambda(x_t,x_{t+1}) \frac{\varphi_{t+1}(x_{t+1})}{\varphi_t(x_t)},
\end{align}
for all $t \in \{0,1, \ldots, T\}$. 

Thus, we obtained explicit expressions for the updated transition probabilities \eqref{eq:markovdis}, which may also vary as a function of time. It remains to find a value for $\lambda$ to ensure that 
the last ensemble-path condition in \eqref{eq:discrete-SBP} is satisfied. 
Feasibility is not automatic, since this last condition departs from the standard setting of the Schr\"odinger bridge problem. 
Assuming feasibility (e.g., by arguing on physical grounds that the data represent indeed
such measurements), an optimal value for $\lambda$ exists due to convexity.
Determining the value for the optimal $\lambda$ can proceed through an iterative search, akin to the one in the continuous case (see Section \ref{sec:lagrange}). In this case, this search is faster due to the discrete nature of the problem. Moreover, having time-varying constraints (c.f. \eqref{eq:dyn-constr}) becomes more approachable since the search for $\lambda(t)$ is a finite-dimensional search as well.

\subsection{The Maximum Caliber Problem}\label{sec:MaxCal}
Let us now tackle the more standard problem of Maximum Caliber. Specifically, in analogy to the classical case, we consider the following problem: 
\begin{align*}
   & \min_{P \ll Q}  \sum_{\mathbf x \in \mathcal X^{T+1}} P (\mathbf x) \log \frac{P (\mathbf x)}{Q (\mathbf x)}, \\
   &{\rm s.t. }\sum_{x_0, \cdots, x_T \in \mathcal X} \!\left(\sum_{k=0}^{T-1} j_k(x_k, x_{k+1}) \right)\! P(x_0, x_1 ,\cdots, x_T) = \text{\j}^{\rm exp},
\end{align*}
i.e., without specific boundary constraints.
The optimizer takes the form
\begin{equation*}
     P^\star (\mathbf x) = \frac{e^{- \sum\limits_{k=0}^{T-1} \lambda j_k(x_k, x_{k+1})} Q(\mathbf x)}{\mathcal Z}.
\end{equation*}
Setting
\begin{align*}
    \hat \varphi_0(x_0) = q_0(x_0),\ \mbox{ and }\ 
    \varphi_T(x_T) = \frac{1}{\mathcal Z},
\end{align*}
the optimizer can be written as
\begin{equation*}
     P^\star (\mathbf x) = \frac{\hat\varphi_0(x_0)}{  q_0(x_0)} \varphi_T(x_T)e^{- \sum\limits_{k=0}^{T-1} \lambda j_k(x_k, x_{k+1})} Q(\mathbf x),
\end{equation*}
 with the optimal Markov kernel obtained via equation \eqref{eq:markovdis}, and the one-time marginals given by~\eqref{eq:dis-phiphihat}, where the functions $\varphi_t$ and $\hat\varphi_t$ are obtained through (\ref{eq: fordis},\ref{eq: backdis}).

\subsection{Steady-state}
\label{sec:discrete-ss}

We now return to the maximum caliber problem with stationary conditions.
Specifically, we consider a steady-state prior $Q$ and a  current constraint of the form 
$$
\sum_{x_0,x_1\in\mathcal X} j(x_0,x_1) P(x_0,x_1)=\text{\j}^{\rm exp},
$$
and we search for the most likely steady-state law $P$ that satisfies this constraint. Here, it is enough to consider a path $\mathbf x=(x_0,x_1)$ with only two successive points in time. Once again, without loss of generality, we assume a scalar such constraint -- a vector-valued constraint can be dealt with similarly.
The problem can be written in the form of a Schr\"odinger bridge problem \eqref{eq:discrete-SBP} with identical endpoint constraints $p_0^{\rm exp}(x)=p_1^{\rm exp}(x)=p(x)$ that need to be determined.
Thus, $p(x)$ is free (sums to $1$) and has to be optimized for.

Therefore, the optimal law has the form \eqref{eq: optdis}, and must be such that \eqref{eq:constr-phi} is satisfied for $p_0^{\rm exp}(x_0)=p(x_0)$ and $p_1^{\rm exp}(x_1)=p(x_1)$. Namely, for any $x_0,\, x_1$
\begin{align*} \hat\varphi_0(x_0)\varphi_0(x_0)=p(x_0) \ \mbox{ and }\ \hat\varphi_1(x_1)\varphi_1(x_1)=p(x_1),
\end{align*}
where we only need to consider one step,
\begin{subequations}\label{eq:Perron}
   \begin{align}
    &\hat \varphi_{1}(x_{1}) = \sum_{x_0}  q_{x_{1} \mid x_0} \Lambda(x_0,x_{1}) \hat \varphi_0(x_0) , \\
 &   \varphi_0(x_0) = \sum_{x_{1}}  q_{x_{1} \mid x_{0}} \Lambda(x_0,x_{1}). \varphi_{1}(x_{1}) 
\end{align} 
\end{subequations}
This implies that $\hat\varphi_0(x)\varphi_0(x)=\hat\varphi_1(x)\varphi_1(x).$ We choose
\begin{equation*}
\hat\varphi_1(x)=\eta\hat\varphi_0(x)\ \mbox{ and }\ \varphi_1(x)=\frac{1}{\eta}\varphi_0(x).
\end{equation*}
Then, $\varphi$ and $\hat\varphi$ are uniquely fixed by \eqref{eq:Perron}. To see this, note that $\hat\varphi_0$ and $\varphi_1$ are the right and left Perron-Frobenius eigenvectors of the matrix 
\[
[q_{x_{1} \mid x_{0}} \Lambda(x_0,x_{1})]_{x_0,x_{1}\in\mathcal X},
\]
respectively, with eigenvalue $\eta$; see \cite{horn2012matrix} for the Perron-Frobenius theory that ensures a unique maximal real and positive eigenvalue $\eta$, and \cite{chen2016robust} for an analogous framework to ours in the context of networks.
Evidently, 
$$
p(x)=\hat\varphi_t(x)\varphi_t(x),
$$
is the same for all $t$, and the updated  time-homogeneous Markov kernel is given by
$$
   p_{x_{1} \mid x_0} = q _{x_{1} \mid x_0} \Lambda(x_0,x_{1}) \frac{\varphi_{0}(x_{1})}{\eta\varphi_0(x_0)}.
$$
Thus, we recover previous steady-state results
\cite{dixit2015stationary,dixit2018communication}.

\section{Examples}
\label{sec:examples}

\subsection{Schrödinger bridges with currents:\\\ \ \ \ \  \ Quadratic potential with hidden DoF}
\label{sec:ex1}
Let us consider a two-dimensional colloidal particle subject to a quadratic potential (exerted for instance by optical tweezers). This quadratic potential has two contributions. 
The first comes from a force applied along the first degree of freedom that we have control over, while the second comes from an underlying unknown potential, which we want to estimate, affecting a degree of freedom that remains ``hidden." 
The force under our control extracts an amount of work from the system that we can measure and provides the data for our modeling and estimation problem.

Specifically, we consider the following potential energy
$$
U(t,x)=\frac{1}{2}x'(K^{\rm prior}+C(t))x,
$$
where $x\in \mathbb R^2$, and 
$$
K^{\rm prior}=\left[\begin{array}{cc}
    3  & 1\\
 1  &  2
\end{array}\right], \quad C(t)=\left[\begin{array}{cc}
    \cos(t) & 0 \\
   0  &  0
\end{array}\right]
$$
represent the intensity of the unknown underlying potential we want to estimate, and the control force, respectively.
Hence, the prior dynamics follow
\begin{align*}
    dX_t=-\frac{1}{\gamma}(K^{\rm prior}+C(t))X_tdt+\sqrt{\frac{2k_BT}{\gamma}}dW_t,
\end{align*}
where $\{W_t\}$ is a two-dimensional Brownian motion, $\gamma$ represents the friction coefficient, $k_B$ the Boltzmann constant and $T\in\mathbb R$ the temperature of the heat bath. The work extracted from the system over a time interval $[0,T]$ by our control force is given by
\begin{align}\nonumber
    W&=-\frac12\int_0^T\int x'\dot C(t)xp(t,x)dx dt\\&=\frac12 \int_0^T \Sigma_{11}(t)\sin(t)dt,\label{eq:work}
\end{align}
where $\Sigma_{11}$ denotes the one-one component of the covariance matrix $\Sigma=\int xx'pdx$, and $\dot{C}$ denotes the time derivative of $C$. In this first example, we measure 
work extraction  $W^{\rm exp}$ along a time interval $[0,T]$, when our system starts and ends at Gaussian distributions with zero-mean and covariances
$$
\Sigma_0^{\rm exp}=\left[\begin{array}{cc}
    1  & 0.75\\
 0.75  &  1.5
\end{array}\right], \mbox{ and }
\Sigma_T^{\rm exp}=\left[\begin{array}{cc}
    0.2 &   -0.01    \\
 -0.01  & 0.5
\end{array}\right],
$$
respectively. However, the measured data is not consistent with the prior dynamics and we are interested in the most likely path that the ensemble of particles took to produce the measurements we obtained. The resulting dynamics allow inference of the underlying potential affecting the hidden degree of freedom.

To this end, we apply the formalism developed in \ref{sec:BC} to the particular dynamics at hand. Due to the quadratic nature of the potential and the Gaussian endpoints,  $\varphi$, and $\hat\varphi$ can be taken as an ansatz to be of the form
$$
\varphi(t,x)=N(t)e^{-\frac12 x'A(t)x}\mbox{ and }\hat\varphi(t,x)=M(t)e^{-\frac12 x'B(t)x}.
$$
Then, due to \eqref{eq:phi} and \eqref{eq:hat-dyn}, $A(t)$ and $B(t)$ must satisfy
\begin{align*}
\dot A&=(K^{\rm prior}+C)'A+A'(K^{\rm prior}+C)+\sigma^2A'A+\lambda \dot C,\\
\dot B&=(K^{\rm prior}+C)'B+B'(K^{\rm prior}+C)-\sigma^2B'B-\lambda \dot C,
\end{align*}
where we have used the fact that $j(t,x)=-\frac12x'\partial_tC(t)x$, and have set $\sigma^2=\frac{2k_BT}{\gamma}$. The boundary conditions must be such that
$$
\big(\Sigma_0^{\rm exp}\big)^{-1}=A(0)+B(0) \mbox{ and } \big(\Sigma_T^{\rm exp}\big)^{-1}=A(T)+B(T). 
$$
Moreover, the optimal $\lambda$ must lead to
\begin{equation}\label{eq:work-constr}
    -\frac12\int_0^T\int x'\dot{C}(t)x\varphi(t,x)\hat\varphi(t,x)dxdt=W^{\rm exp}.
\end{equation}

\begin{figure}[t]
    \centering
\includegraphics[width=0.48325\textwidth,trim={2.125cm 8.75cm 1.1cm 0.55cm},clip]{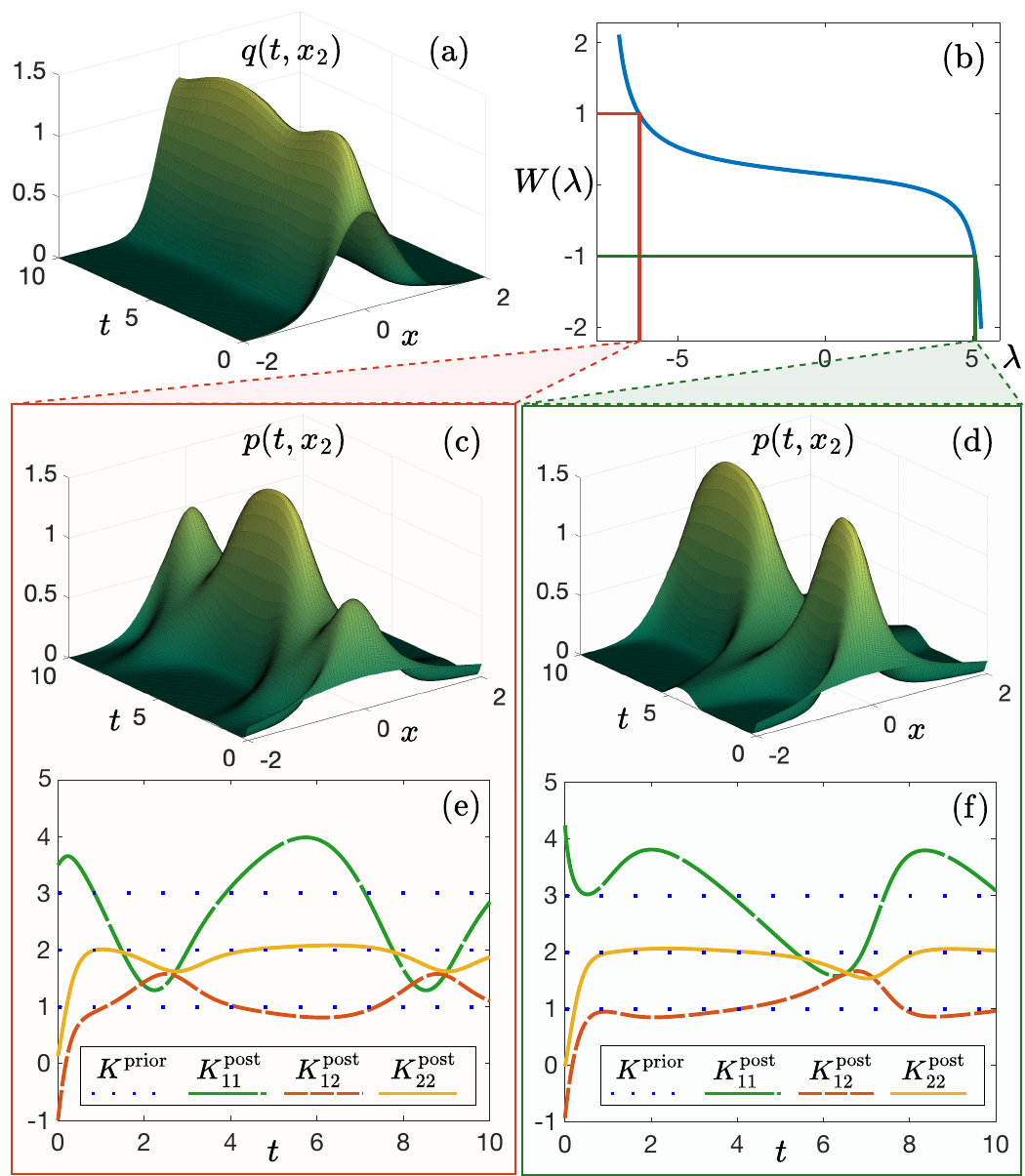}
    \caption{Prior and posterior distributions along the second degree of freedom $x_2$, together with their associated potentials. (a) Prior one-time marginals along $x_2$. (b) Work extracted for different values of $\lambda$. Blow-up figures correspond to the values of $\lambda$ that satisfy $W(\lambda)=1$ and $W(\lambda)=-1$, respectively. (c) Posterior one-time marginals along $x_2$ for $W(\lambda)=1$.
    (d)~Posterior one-time marginals along $x_2$ for $W(\lambda)=-1$.
    (e/f)~Entries of the potential intensity matrix $K^{\rm post}$ under which the posterior distribution in (c/d) is typical; the values of entries are drawn (continuous curves) on top of those (dotted lines) corresponding to the prior potential intensity matrix $K^{\rm prior}$. }
    \label{fig:1}
\end{figure}

To find the appropriate boundary conditions $A(T)$, $B(0)$ and the value of the Lagrange multiplier $\lambda$, we first apply a Sinkhorn algorithm for different $\lambda$'s to find the boundary constraints $A(T)$ and $ B(0)$, along with the obtained work $W(\lambda)$ \footnote{We follow this approach for illustrative purposes. In general, it is more efficient to iterate over $\lambda$ using the gradient descent method described in equation \eqref{eq:grad-asc}}. Then, we find the particular $\lambda$ that satisfies equation \eqref{eq:work-constr}. 
The results of this procedure for $W^{\rm exp}=1$ and $W^{\rm exp}=-1$ (the negative denoting work is put \emph{into} the system) are shown in Fig.~\ref{fig:1}. Therein we observe how for the different measured work values, we obtain vastly different posterior distributions~(c,\,d), along with different estimates of the underlying potential intensity (e,\,f)
$$
K^{\rm post}(t)=K^{\rm prior}+\sigma^2A(t).
$$
These results highlight the impact that a measurement that pertains to the first degree of freedom has on estimating one-time marginals along the second (hidden) degree of freedom.
Indeed, the posterior distributions and potentials have approximately opposite phases in these two cases. 
This can be understood by observing that to obtain positive work output the phase of $\Sigma_{11}(t)$ must approximately match that of $\sin(t)$ (c.f. \eqref{eq:work}), while the phase must be opposite if negative work output is obtained. Moreover, it is also seen that to obtain positive work output, $\lambda$ has to be further away from $0$ ($\lambda=-6.31$) than in the negative work output case  ($\lambda=5.09$). 
This seems reasonable since extracting work in finite time is thermodynamically costly (due to unavoidable dissipation), and hence more constrained than putting work in.

\subsection{Maximum Caliber: a bit erasure experiment}
\label{sec:ex2}

We consider a bit erasure experiment~\cite{landauer1991information}, in which the state of the system starts being uniformly distributed over two possibilities ($0$ and $1$), and is driven towards a specified state (say, being reset to $0$).
This kind of experiment can be realized through a trapped colloidal particle \cite{berut2015information} or nanomagnetic memory bits \cite{hong2016experimental}. 
We consider an experiment in which the only available information is the success rate in erasing the bit, that is, the probability mass corresponding to state 0 at the final time. 
Assuming that the observation does not match the prior dynamics, we adapt the Maximum Caliber formalism in \ref{sec:MaxCal} to account for this discrepancy.

\begin{figure}[tb]
    \centering
\includegraphics[width=0.48325\textwidth,trim={0.6cm 12.67cm 13.825cm 0.47cm},clip]{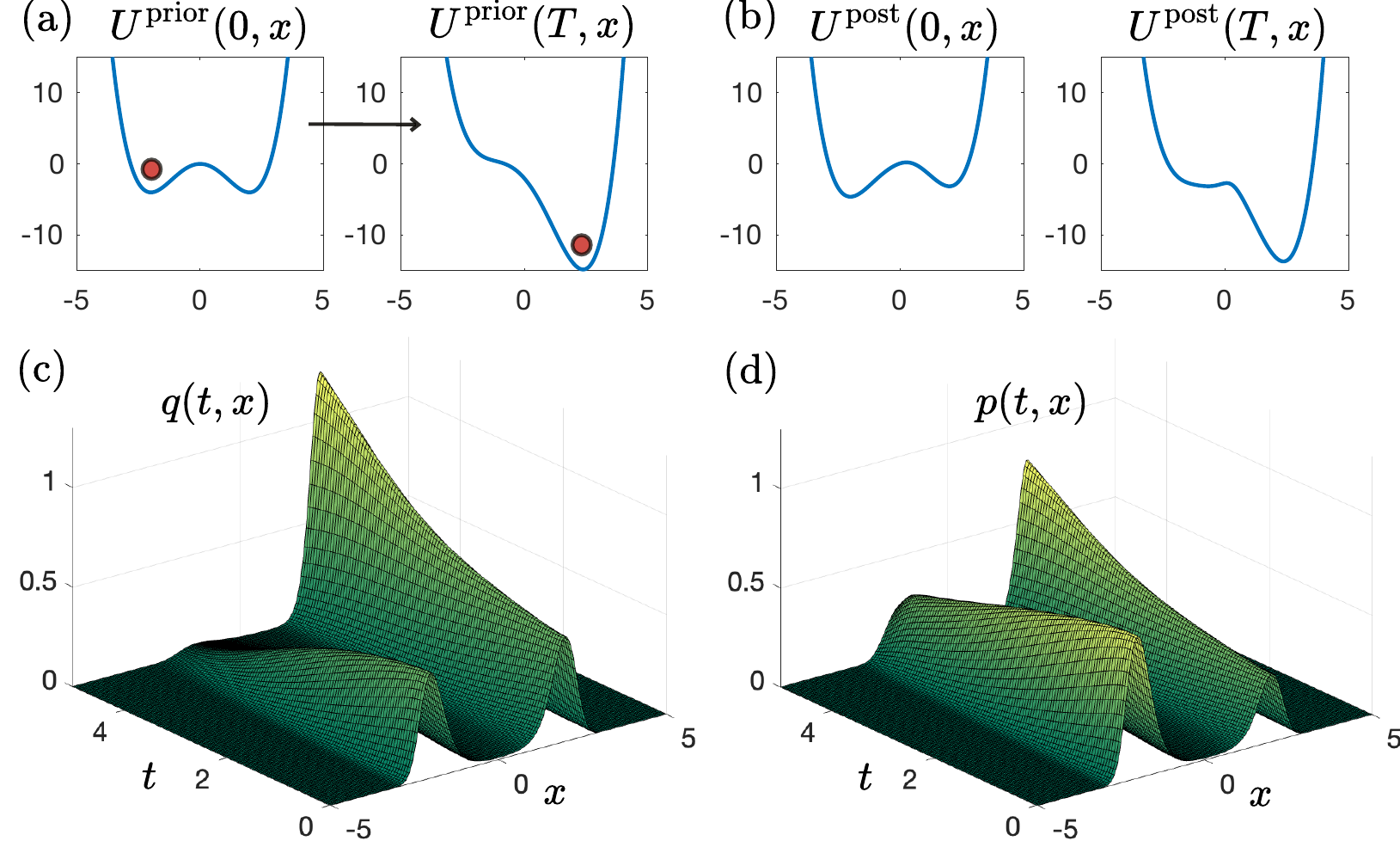}
    \caption{Prior and posterior distributions of a bit erasure experiment, along with the corresponding potential energies plotted at initial and final times for simplicity. (a) Prior potential energy at initial and final times; the arrow designates the motion of a typical particle being reset to 0. (b) Posterior potential energy at initial and final times. 
    (c) Prior density flow that begins at equilibrium with particles distributed equally over the two sides of a double well potential, and ending with $92.6\%$  of the particles in the right well. 
    (d) Most likely flow of the probability distribution of particles that end with only $60\%$ in the right well.}
    \label{fig:2}
\end{figure}

Following \cite{landauer1991information,berut2015information}, we assume the prior dynamics of our colloidal particle undergoing bit erasure to be
$$
dX_t=-\frac{1}{\gamma}\nabla U^{\rm prior}(t,X_t)dt+\sqrt{\frac{2k_B T}{\gamma}}dW_t,
$$
where $X_t\in\mathbb R$, $\{W_t\}$ is the one-dimensional Brownian motion, and the potential function $U(t,X_t)$ is given by
$$
U^{\rm prior}(t,x)=\frac{1}{4}x^4-2(x+0.2t)^2,
$$
with $t\in[0,T],$ where $T=5.$ The experiment starts at an equilibrium distribution in which the particles are in the left and right wells with equal probability. ``Tilting" of the potential steers the particles towards the right well (see Figure \ref{fig:2}(a)). At the end of the experiment, we expect the bit to be erased, that is, the probability of the particle sitting on the right well to be close to 1 (specifically, 0.926). 
However, we perform the bit erasure experiment and we observe that at the end ($t=T$) the bit is not perfectly erased; the probability of erasure is measured to be
$$
\int_0^\infty p(T,x)dx=\text{\j}^{\rm exp}=0.6.
$$
What is the most likely evolution of the probability density of this bit erasure experiment? What is the potential landscape under which this evolution is typical?

To solve this problem we make use of the framework presented in \ref{sec:MaxCal}, with a slight modification. In this case, our current constraint reads
$$
\int_{\mathbb R} j(T,x_T)p(T,x_T)dx_T=0.6,
$$
with $j(T,x_T)=H(x_T)$ where $H$ is the Heaviside function. Therefore, the optimal posterior distribution must be of the form (c.f. \eqref{eq:RN2})
$$
P^\star=\frac{e^{-\lambda j(T,X_T)}Q}{\mathcal Z}.
$$
This leads to the one-time marginals 
$$
p(t,x)=\varphi(t,x)\hat\varphi(t,x),
$$
with
 $\varphi(t,x),\,\hat\varphi(t,x)$ satisfying
    \begin{align}\label{eq:endp}
           \partial_t\varphi&=-\alpha\nabla\varphi-\frac12\sigma^2\Delta\varphi,\ \varphi(T,x)=\frac{1}{\mathcal Z}e^{-\lambda H(x)}\\\nonumber
           \partial_t\hat\varphi&=-\nabla\cdot(\alpha \hat\varphi)\!+\!\frac12\Delta (\sigma^2\hat\varphi),\ \hat\varphi(0,x)=q_0(x).
    \end{align}

We remark that in this case, $\hat\varphi$ evolves according to the prior Fokker-Planck equation, with the prior initial condition. Thus, $\hat\varphi(t,x)=q(t,x)$. Moreover, in this example, it is not necessary to search for the optimal $\lambda$, since it is explicitly fixed by our constraint
$$
\int_0^\infty p(T,x_T)dx_T=\frac{e^{-\lambda}}{\mathcal Z}\int_0^\infty q(T,x_T)dx_T=0.6,
$$
implying that 
$$
\frac{e^{-\lambda}}{\mathcal Z}=\frac{0.6}{0.926}=0.648.
$$
Since all one-time marginals must be normalized,
$$
\int_{\mathbb R}p(T,x)dx=\frac{1}{\mathcal Z}\int^0_{-\infty}q(T,x)dx+0.6=1,
$$
from where we can obtain $\mathcal Z$ to uniquely fix the endpoint constraint in \eqref{eq:endp}.

Following this procedure, we have obtained the optimal posterior distribution, along with the associated estimate of the potential landscape
$$
U^{\rm post}(t,x)=U^{\rm prior}(t,x)-\frac{2k_BT}{\gamma}\log\varphi(t,x),
$$
 which are depicted in Figure \ref{fig:2}. We observe how the constraint on the endpoint populations smoothly affects the whole trajectory of particles, including the initial distribution and potential.

\subsection{Steady-state: energy landscapes\\ for protein folding}\label{ex:last}

\begin{figure}[tb]
    \centering
\includegraphics[width=0.48325\textwidth,trim={1.4cm 10.32cm 2.6cm 0.075cm},clip]{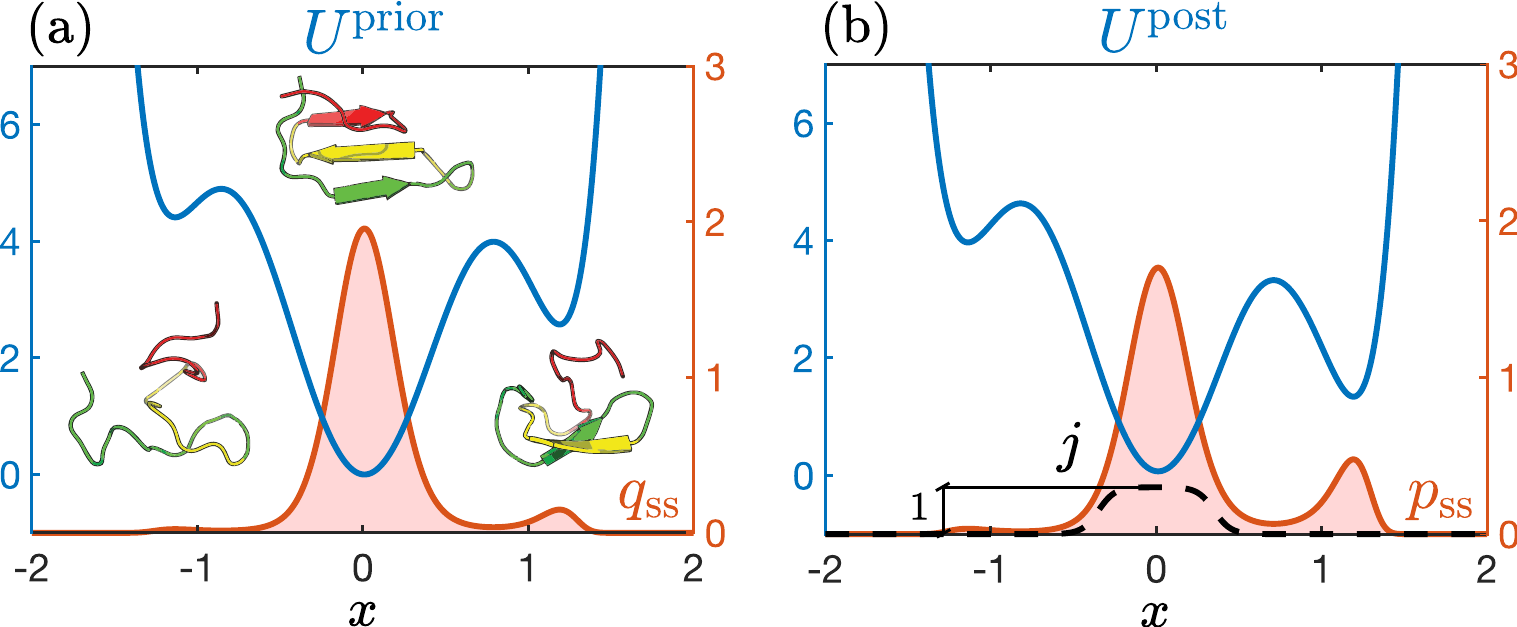}
    \caption{(a) Prior potential and the associated steady-state distribution; schematics of the conformational state of the protein corresponding to the denatured (unfolded), native, and molten globule states are depicted (left to right). (b) Posterior potential and steady-state distribution (solid curves); the dashed curve represents the shape of $j(x)$ as a smooth indicator function on the native state.
    }
    \label{fig:3}
\end{figure}

As a last example, let us consider protein folding kinetics \cite{dill1995principles}. A popular model to describe such kinetics is based on energy landscapes on the conformation space of the protein \cite{bryngelson1995funnels,shea2001folding} (see Figure \ref{fig:3}(a) for an example).
Denatured (unfolded) proteins move across those (typically steady) energy landscapes through some noisy dynamics~\cite{veitshans1997protein}, eventually reaching the native (folded) state. Specifically, let $X_t\in\mathbb R$ denote the conformation state of the protein, which we assume to evolve according to~\cite{yang2006effective,bechhoefer2015hidden}
\begin{equation}
    dX_t=-\nabla U^{\rm prior}(X_t)dt+\sigma dW_t,
\end{equation}
where $U^{\rm prior}(x)$ is the energy landscape depicted in Figure \ref{fig:3}(a).
The considered landscape has two metastable states corresponding to the unfolded state (left well) and a molten globule state (right well). The stable state (middle well) corresponds to the folded state in which the protein is biologically functional. An ensemble of such proteins reaches a steady-state $q_{\rm ss}$, in which $82.5\%$ of the proteins are at the native state.

Assume we measure the concentration of proteins at steady-state in the native state to be less than what we expected from $q_{\rm ss}$. That is, we measure
\begin{equation}\label{eq:protein-constr}
    \int_{\mathbb R^n}j(x)p_{\rm ss}(x)dx=0.72,
\end{equation}
where $j(x)$ is the smoothed indicator function of the native state depicted in Figure \ref{fig:3}(b). The most likely steady-state that led to that measurement, along with the energy landscape that generates it, can be found by solving for $\varphi(x)$ and $\hat\varphi(x)$ in \eqref{eq:ss-cont}. Then, as usual,
$$
p_{\rm ss}(x)=\varphi(x)\hat\varphi(x)$$
and $$ U^{\rm post}(x)=U^{\rm prior}(x)-\sigma^2\log\varphi(x),
$$
up to a constant. Iteration over values of $\lambda$ is necessary to find the one that satisfies the constraint \eqref{eq:protein-constr}. The corresponding optimal solution is portrayed in Figure \ref{fig:3}(b). As noted earlier, the obtained solution is not equivalent to the naive entropy-maximizing steady-state distribution.

The dynamical counterpart of this theory may be used to determine dynamical energy landscapes of coupled binding and protein conformational change \cite{tsai1999folding,okazaki2008dynamic}.
Indeed, inferring energy landscapes is invaluable in identifying and designing novel protein-binding ligands for drug discovery \cite{sledz2018protein,dickson2018mapping}.

\section{Conclusions}

The goal of this work has been to link and further develop certain classical viewpoints for solving inverse problems, seeking the most likely explanation of experimentally collected measurements. Historically, the main ideas go back to the dawn of statistical mechanics at the beginning of the 19th century. Our theme however begins with a proposal by Erwin Schr\"odinger, the so-called Schr\"odinger bridge, that seeks an update to a prior law so as to restore consistency between measured marginal distributions at different points in time and the given prior law. 
The celebrated Maximum Entropy and Maximum Caliber methods represent extensions of this basic idea, and this is the message that we hope to convey.

On the practical side, the significance of Schr\"odinger's dictum is that the most likely explanation gives rise to a new law, and as a consequence, to an updated dynamical model. For many problems in physics and biology, dynamics are intimately connected to an energy landscape that steers stochastic systems between measurements. 
Estimating potential landscapes, that dictate underlying microscopic physics (e.g. material properties), is of fundamental relevance across science.
Indeed,  the subtle relationship between 
microscopic phenomena -- such as chemical interactions, molecular mechanics, or transition rates -- and macroscopic and thermodynamic properties is of key importance in chemistry, biology, physics, and material science. 

In biology for instance, distinctive energy landscapes are responsible for the workings of F0F1-ATP synthase rotary motors  \cite{wang1998energy,mukherjee2011electrostatic}, as well as 
the sliding of kinesin molecules across microtubule networks \cite{kolomeisky2007molecular}. In material science and chemistry similar examples abound. 
In our last example, Section \ref{ex:last}, even if at a rudimentary level, we have seen how the shape of energy landscapes can provide 
descriptions of dynamics along conformation states 
that capture key thermodynamic and kinetic quantities. 
The practical significance of understanding energy landscapes cannot be underestimated, as it drives protein engineering and the design of molecules and materials with specified macroscopic properties \cite{panettieri2018energy,zhang2020assembly}.

On this broad template, the paradigm of Maximum Caliber brings yet another dimension to the type of measurements that one may consider, ensemble-path measurements. As we have seen in the body of the paper, these can be seamlessly treated within the frame of Schr\"odinger bridges.
Ensemble-path measurements are especially interesting as they may represent work, currents, and other physical quantities of timely importance in the rapidly developing technological front that peers into the microscopic world.

In this work, we have presented an integrated view of Schrödinger bridges and Maximum Caliber to provide optimal estimates of both ensemble dynamics and (time-varying) potential landscapes. 
The framework is computationally tractable
with an elegant structure that dates back to Schrödinger. 
We see as items for future research the following. First, 
the fact that, in general, we obtain time-varying dynamics
may be impractical in cases where potential landscapes 
are ``known" to be constant. Thus, how to come up with \emph{time-invariant} landscapes (or other parameters), without assuming that the system is at steady-state, is of great interest.
Second, we note that 
the presented theory can only account for measured currents $j(t,x)$ that are independent of the model, i.e., independent of both the ensemble distribution $p(t,x)$ and the drift $\beta(t,x).$ Therefore, 
certain currents of interest, such as heat or entropy, cannot be accounted for directly.
An adaptation of the formalism
to account for such ``coupled" currents 
would be highly valued. 
Finally, our theory opens up the study of a new Optimal Mass Transport problem in which currents are fixed. Specifically, it is of interest to explore whether the Optimal Mass Transport problem with currents gives rise to a meaningful distance on the space of probabilities; a useful metric in which gradient flows are envisioned to provide optimal models in the sense of Maximum Caliber.


\black

\begin{acknowledgments}
O.M.M. gratefully acknowledges Artemy Kolchinsky for pointing to the relationship between Schrödinger bridges with currents and MaxCal. Protein sketches in Figure \ref{fig:3}(a) have been adapted from scistyle.com.
\end{acknowledgments}

\appendix

\bibliography{references}

\end{document}